\pdfoutput=1

\documentclass[preprint,preprintnumbers,amsmath,amssymb,floatfix,endfloats*]{revtex4}

\usepackage{graphicx}
\usepackage{dcolumn}
\usepackage{bm}
\usepackage{amsfonts}

\UseRawInputEncoding

\DeclareMathAlphabet{\mathsfsl}{OT1}{cmr}{bx}{it}
\begin{document}
%
\title{Fatigue behavior of Cu-Zr metallic glasses under cyclic loading}
\author{Nikolai V. Priezjev$^{1,2}$}
\affiliation{$^{1}$Department of Civil and Environmental
Engineering, Howard University, Washington, D.C. 20059}
\affiliation{$^{2}$Department of Mechanical and Materials
Engineering, Wright State University, Dayton, OH 45435}
\date{\today}
\begin{abstract}

The effect of oscillatory shear deformation on the fatigue life,
yielding transition, and flow localization in metallic glasses is
investigated using molecular dynamics simulations. We study a
well-annealed Cu-Zr amorphous alloy subjected to periodic shear at
room temperature. We find that upon loading for hundreds of cycles
at strain amplitudes just below a critical value, the potential
energy at zero strain remains nearly constant and plastic events are
highly localized. By contrast, at strain amplitudes above the
critical point, the plastic deformation is gradually accumulated
upon continued loading until the yielding transition and the
formation of a shear band across the entire system. Interestingly,
when the strain amplitude approaches the critical value from above,
the number of cycles to failure increases as a power-law function,
which is consistent with the previous results on binary
Lennard-Jones glasses.

\vskip 0.5in


\end{abstract}

\maketitle

\section{Introduction}

Understanding structure-property-function relationships in metallic
glasses, which are known to possess exceptionally high strength and
large elastic strain limit, is important for numerous structural
applications~\cite{Kruzic16,Qiao19}. Under cyclic loading at
sufficiently large stress or strain amplitudes, such amorphous
alloys ultimately fail via the formation of nanoscale shear bands
and propagation of microscale cracks~\cite{Sha2020,Liaw18,Menzel06}.
On the microscopic level, an elementary plastic event in disordered
alloys involves a swift collective rearrangement of a few tens of
atoms or a shear transformation~\cite{Spaepen77,Argon79}. Using a
mean field elastoplastic model with a distribution of local yield
barriers, it was recently demonstrated that fatigue life of athermal
amorphous solids follows a power-law divergence when the yielding
point is approached from above~\cite{Sollich22}. Furthermore, within
two models of elastoplastic rheology, it was found that fatigue in
amorphous materials proceeds via slow accumulation of low levels of
damage and it is followed by a sudden failure and shear band
formation after a number of cycles that, in turn, depend on the
strain amplitude, working temperature, and degree of annealing prior
to loading~\cite{Fielding22}. Despite recent advances, however, the
exact mechanism of fatigue failure in metallic glasses and an
accurate prediction of the fatigue life remain not fully understood.

\vskip 0.05in

In the last few years, the effect of cyclic loading on rejuvenation,
relaxation, and yielding phenomena in amorphous alloys was
extensively studied using particle-based
simulations~\cite{Priezjev13,Reichhardt13,Sastry13,IdoNature15,Shi15,
GaoNano15,Priezjev16,Kawasaki16,Priezjev16a,Keblinski17,Sastry17,
Priezjev17,Priezjev18,Priezjev18a,Sastry19band,
NVP18strload,She19,PriezSHALT19,Priez20ba,KawBer20,NVP20altY,Bauchy20,
WangYang20,Priez20delay,BhaSastry21,Priez21var,Notch21,PriezCMS21,
Peng22,Chan22,PriezJNCS22,Barrat23,PriezCMS23}. In particular, it
was found that after a number of transient cycles, athermal glasses
undergo strain-induced reversible deformation where trajectories of
atoms become identical during one or more
periods~\cite{Reichhardt13,IdoNature15}. Whereas poorly annealed
glasses tend to relax when subjected to small-amplitude cyclic
shear~\cite{Sastry13,Priezjev18,PriezSHALT19,Sastry19band,KawBer20},
well-annealed amorphous alloys can be rejuvenated during a number of
cycles before yielding at strain amplitudes above a critical
point~\cite{Priez21var,Priez20delay,PriezCMS23}. The yielding
transition in periodically deformed glasses is typically associated
with the formation of shear bands initiated either at open
boundaries~\cite{GaoNano15,Shi15,She19,Notch21,Chan22} or in the
bulk of the computational domain in case of periodic boundary
conditions~\cite{Sastry17,Priezjev17,Priezjev18a,
Sastry19band,Priez20ba,Priez20delay,PriezCMS21,PriezJNCS22,
PriezCMS23}. In the vicinity of a critical strain amplitude, the
number of cycles to reach the yielding transition depends on
temperature~\cite{Priez20delay}, preparation
history~\cite{Sastry17,Priez20ba,PriezCMS21,PriezJNCS22}, strain
amplitude~\cite{GaoNano15, Sastry17, Priezjev17, Priez20delay},
deformation protocol~\cite{NVP20altY}, and loading
frequency~\cite{GaoNano15}. Notably, it was recently demonstrated
that fatigue failure in well-annealed binary Lennard-Jones (LJ)
glasses occurs after a number of cycles that increases as a
power-law function when the strain amplitude approaches a critical
value from above~\cite{PriezCMS23}. In spite of significant
progress, however, the role of loading protocol, processing history,
and interaction potentials on the fatigue life and yielding
transition in disordered solids is still being actively
investigated.

\vskip 0.05in

In this paper, we report the results of molecular dynamics (MD)
simulations of a Cu-Zr metallic glass subjected to oscillatory shear
deformation at strain amplitudes near a critical value. We consider
a well-annealed sample prepared by cooling at a slow rate from the
liquid state to room temperature. It will be shown that after a
certain number of shear cycles, the glass undergoes a yielding
transition characterized by the formation of a system-spanning shear
band and a steep increase of the potential energy. We find that upon
reducing strain amplitude towards a critical value, the number of
cycles to reach the yielding transition increases as a power-law
function, which is in agreement with simulation results for
low-temperature binary LJ glasses.

\vskip 0.05in

The rest of this paper proceeds as follows. In the next section, we
describe the preparation procedure and the deformation protocol. The
results for the potential energy, shear stress, nonaffine
displacements, and atomic configurations are presented in
section\,\ref{sec:Results}. A brief summary is given in the last
section.

\section{Molecular dynamics simulations}
\label{sec:MD_Model}

In this study, the $\text{Cu}_{50}\text{Zr}_{50}$ metallic glass was
simulated using the embedded atom method (EAM)
potentials~\cite{CuZrEAM09,Ma11} with the time step $\triangle t =
1.0\,\text{fs}$. The total number of Cu and Zr atoms is $60\,000$.
The system was initially placed into a cubic box and equilibrated at
temperature of $2000\,\text{K}$ and zero pressure. Following the
cooling protocol described by Fan and Ma~\cite{FanMa21}, the liquid
was first cooled to $1500\,\text{K}$ at the rate of
$10^{13}\,\text{K/s}$, then to $1000\,\text{K}$ at
$10^{12}\,\text{K/s}$, and finally to $300\,\text{K}$ at
$10^{10}\,\text{K/s}$. Hence, the effective cooling rate across the
glass transition temperature ($T_g\thicksim675\,\text{K}$) was
$10^{10}\,\text{K/s}$. This preparation procedure was carried out in
the NPT ensemble using the Nos\'{e}-Hoover thermostat, zero external
pressure, and periodic boundary conditions~\cite{Lammps}.

\vskip 0.05in


After cooling to $300\,\text{K}$, the linear size of a cubic box was
fixed to $101.8\,\AA$, and the glass was subjected to oscillatory
shear deformation along the $xz$ plane at constant volume, as
follows:
\begin{equation}
\gamma_{xz}(t)=\gamma_0\,\text{sin}(2\pi t/T ),
\label{Eq:shear}
\end{equation}
where $T=1.0\,\text{ns}$ ($10^6$ MD time steps) is the oscillation
period and $\gamma_0$ is the strain amplitude. The MD simulations
were performed for strain amplitudes in the range $0.055 \leqslant
\gamma_0 \leqslant 0.064$. The potential energy, shear stress, and
atomic configurations at the end of each cycle were periodically
saved for post-processing analysis. Due to computational
limitations, the data were collected only for one realization of
disorder.

\section{Results}
\label{sec:Results}


It is generally realized that mechanical properties of metallic
glasses strongly depend, among other factors, on preparation
history~\cite{Egami13,Greer16}. Thus, sufficiently slowly cooled
glasses settle at lower energy states and upon start-up continuous
deformation tend to exhibit a pronounced yielding peak
characteristic of a brittle response, whereas more rapidly cooled
glasses typically deform in a ductile manner~\cite{Hufnagel16}.
Moreover, it was recently found that well-annealed binary glasses
under cyclic loading yield at strain amplitudes smaller than the
location of the yielding peak~\cite{Sastry17,Priez20ba}. When a
critical value is approached from above, the number of cycles until
the yielding transition increases as a power-law function for binary
LJ glasses at a temperature well below the glass transition
temperature~\cite{PriezCMS23}. In the present paper, these results
are extended to the case of well-annealed
$\text{Cu}_{50}\text{Zr}_{50}$ metallic glasses at room temperature.
We also comment that test runs of more rapidly cooled
$\text{Cu}_{50}\text{Zr}_{50}$ metallic glasses (with the effective
cooling rate of $10^{12}\,\text{K/s}$) have shown that cyclic
loading near the critical strain amplitude induces significant
structural relaxation and a rapid decay of the potential energy
during a number of transient cycles, similar to the results reported
for poorly annealed LJ glasses~\cite{NVP20altY}.

\vskip 0.05in

\begin{figure}[t]
\includegraphics[width=12.0cm,angle=0]{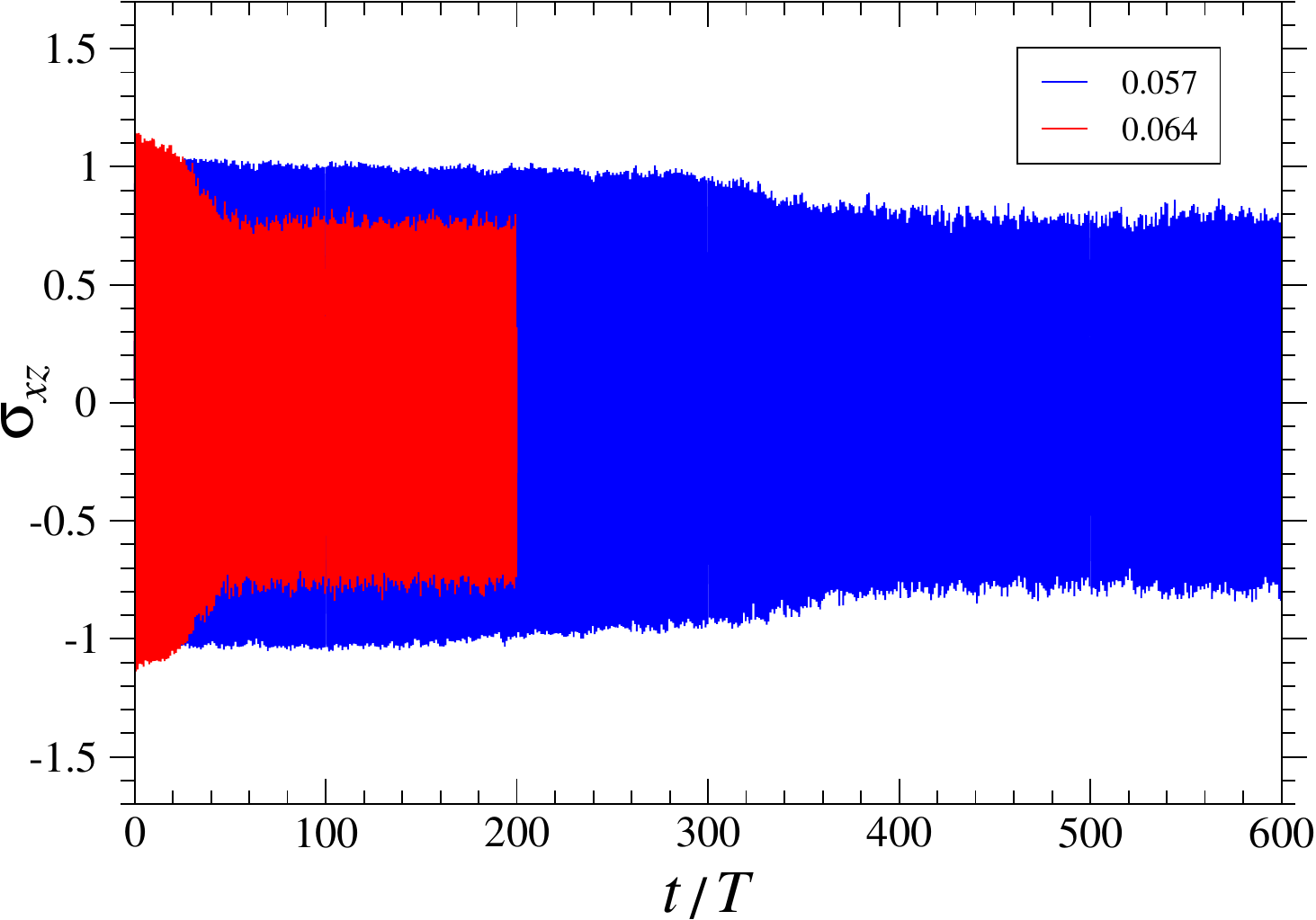}
\caption{(Color online) The time dependence of shear stress,
$\sigma_{xz}$ (in units of GPa), for strain amplitudes
$\gamma_0=0.057$ (blue curve) and $\gamma_0=0.064$ (red curve). The
period of oscillation is $T=1.0\,\text{ns}$. }
\label{fig:stress_xz_amp057_064}
\end{figure}

We first present the dependence of shear stress along the $xz$ plane
as a function of time in Fig.\,\ref{fig:stress_xz_amp057_064} for
strain amplitudes $\gamma_0=0.057$ and $0.064$. Note that for the
largest strain amplitude $\gamma_0=0.064$ considered in the present
study, the glass gradually yields after about 40 shear cycles,
followed by plastic flow within a shear band upon continued loading.
It can be observed that the number of cycles until the yielding
transition increases significantly when the strain amplitude is
reduced to $\gamma_0=0.057$. Note that the amplitude of stress
oscillations after the yielding transition is nearly the same in
both cases, and it is determined by the maximum stress within a
shear band.  A similar behavior of the shear stress was observed for
low-temperature binary LJ glasses under periodic shear, although the
yielding transition in the previous study was associated with a
sharp drop in stress during only one shear cycle~\cite{PriezCMS23}.

\vskip 0.05in

\begin{figure}[t]
\includegraphics[width=12.0cm,angle=0]{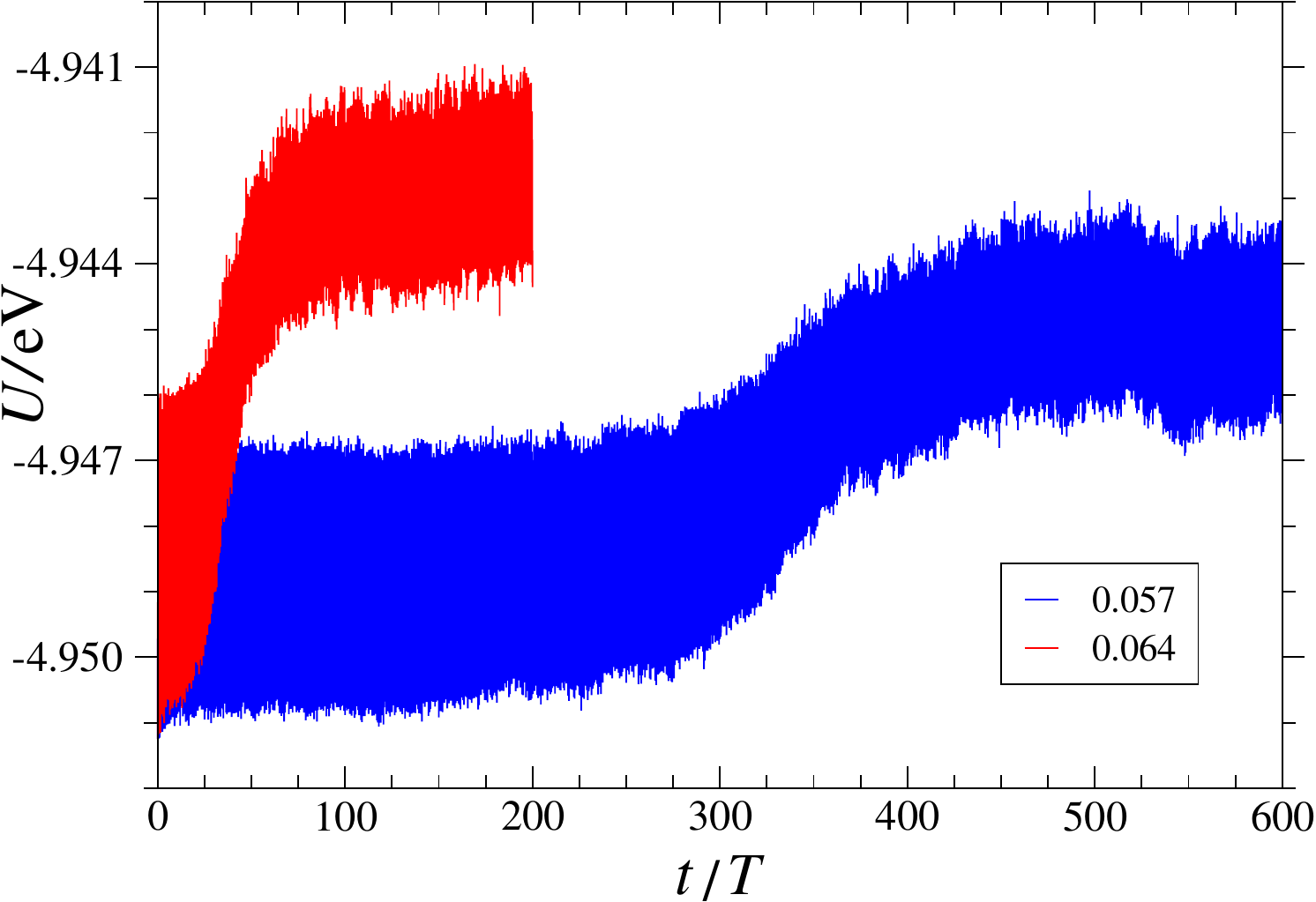}
\caption{(Color online) The variation of the potential energy $U$
(in units of $\text{eV}$ per atom) as a function of time for strain
amplitudes $\gamma_0=0.057$ (blue curve) and $\gamma_0=0.064$ (red
curve). The period of oscillation is $T=1.0\,\text{ns}$.}
\label{fig:poten_T300_r10e10_amps_057_064}
\end{figure}

Next, the variation of the potential energy per atom versus cycle
number is displayed in
Fig.\,\ref{fig:poten_T300_r10e10_amps_057_064} for the same strain
amplitudes, $\gamma_0=0.057$ and $0.064$, as in
Fig.\,\ref{fig:stress_xz_amp057_064}.  Each of the curves in
Fig.\,\ref{fig:poten_T300_r10e10_amps_057_064} clearly show three
stages of deformation under cyclic loading; namely, a solid-like
response, a gradual yielding transition, and plastic flow within a
shear band.  The relatively large scatter of the data is attributed
to thermal fluctuations and finite system size. When loaded at
$\gamma_0=0.064$, the potential energy at the end of each cycle (the
lower envelope) starts to increase abruptly, indicating the
occurrence of localized plastic events that eventually lead to the
formation of a shear band and the corresponding plateau after about
60 cycles. By contrast, the initial stage of accumulation of plastic
rearrangements at the strain amplitude $\gamma_0=0.057$ takes about
300 cycles, and it is followed by a gradual yielding transition
during 150 shear cycles.

\vskip 0.05in

\begin{figure}[t]
\includegraphics[width=12.0cm,angle=0]{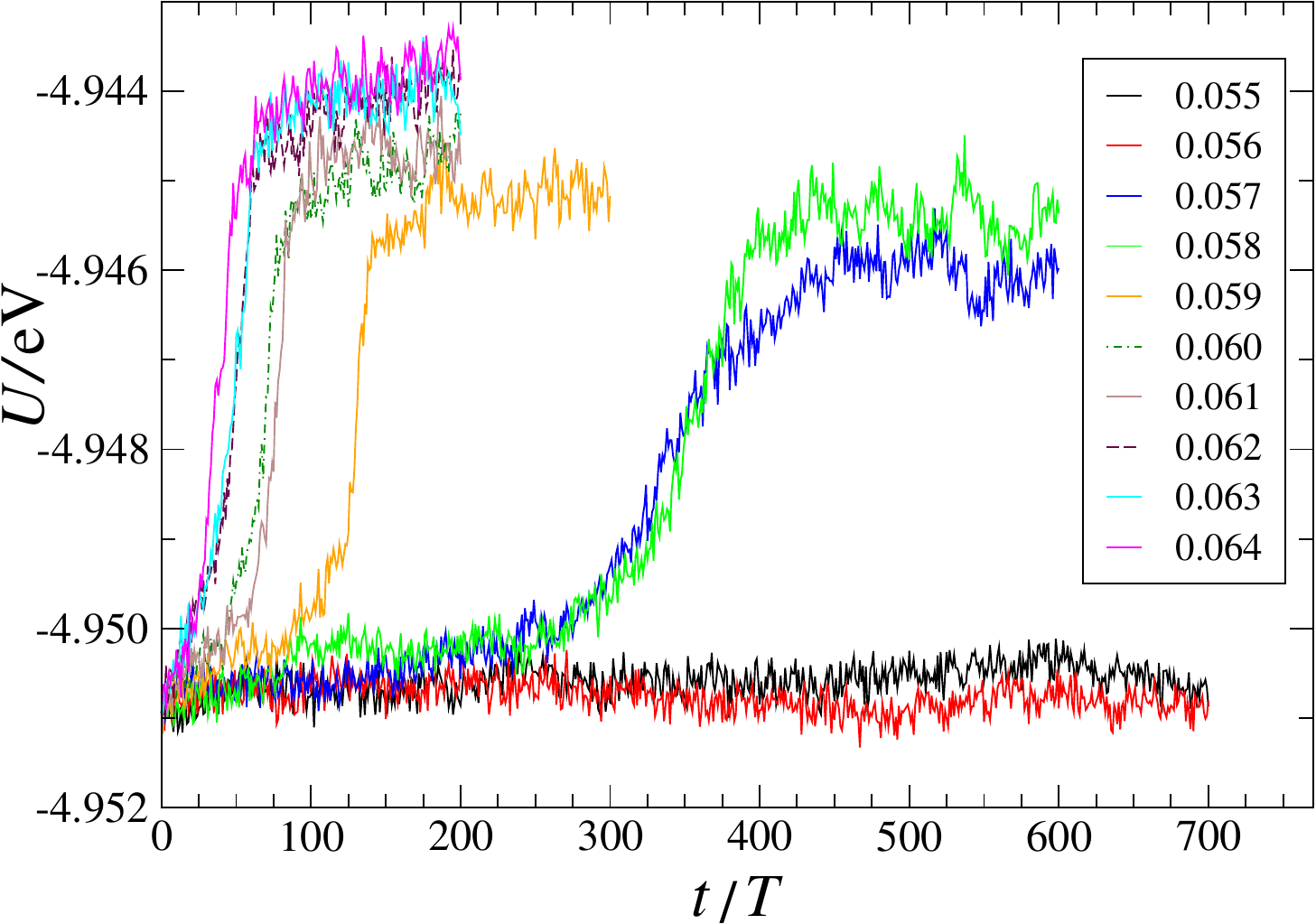}
\caption{(Color online) The potential energy at the end of each
cycle (when $\gamma_{xz}=0$) for strain amplitudes in the range
$0.055 \leqslant \gamma_0 \leqslant 0.064$. The oscillation period
is $T=1.0\,\text{ns}$. }
\label{fig:poten_T300_r10e10_amps}
\end{figure}

A summary of the data for the potential energy after each cycle at
zero strain as a function of the cycle number for strain amplitudes
in the range $0.055 \leqslant \gamma_0 \leqslant 0.064$ is presented
in Fig.\,\ref{fig:poten_T300_r10e10_amps}. It is evident that the
number of cycles until the yielding transition becomes larger when
the glass is loaded at a lower strain amplitude. Notice that the
system did not yield for at least 700 cycles when periodically
deformed at strain amplitudes $\gamma_0=0.055$ and $0.056$.
Therefore, it can be deduced that the critical strain amplitude for
the yielding transition is between $\gamma_0=0.056$ and $0.057$.
Further, the results in Fig.\,\ref{fig:poten_T300_r10e10_amps}
demonstrate that the well-annealed $\text{Cu}_{50}\text{Zr}_{50}$
metallic glass can be rejuvenated via cyclic loading at strain
amplitudes slightly above a critical value, which is similar to
conclusions for binary LJ glasses~\cite{PriezCMS23}. The number of
cycles needed to increase the potential energy by $0.001\,\text{eV}$
per atom varies from about 20 to 300 depending on the strain
amplitude (see Fig.\,\ref{fig:poten_T300_r10e10_amps}).

\vskip 0.05in


It is well known that plastic deformation in disordered solids
involves the so-called nonaffine rearrangements of
atoms~\cite{Falk98}. In particular, the nonaffine measure for
displacement of an atom from $\mathbf{r}_{i}(t)$ to
$\mathbf{r}_{i}(t+\Delta t)$ can be computed using the matrix
$\mathbf{J}_i$, which transforms positions of neighboring atoms and
at the same time minimizes the following expression:
\begin{equation}
D^2(t, \Delta t)=\frac{1}{N_i}\sum_{j=1}^{N_i}\Big\{
\mathbf{r}_{j}(t+\Delta t)-\mathbf{r}_{i}(t+\Delta t)-\mathbf{J}_i
\big[ \mathbf{r}_{j}(t) - \mathbf{r}_{i}(t)  \big] \Big\}^2,
\label{Eq:D2min}
\end{equation}
where the sum is taken over $N_i$ atoms that are initially located
within $4.0\,\AA$ from $\mathbf{r}_{i}(t)$. Irreversible plastic
rearrangements of atoms typically occur when their nonaffine
displacements exceed a cage size~\cite{Priezjev16}. It was
previously shown that the cage size of
$\text{Cu}_{50}\text{Zr}_{50}$ alloy near the glass transition
temperature is about $r_c\approx 0.6\,\AA$~\cite{Douglas19}.

\vskip 0.05in

\begin{figure}[t]
\includegraphics[width=12.0cm,angle=0]{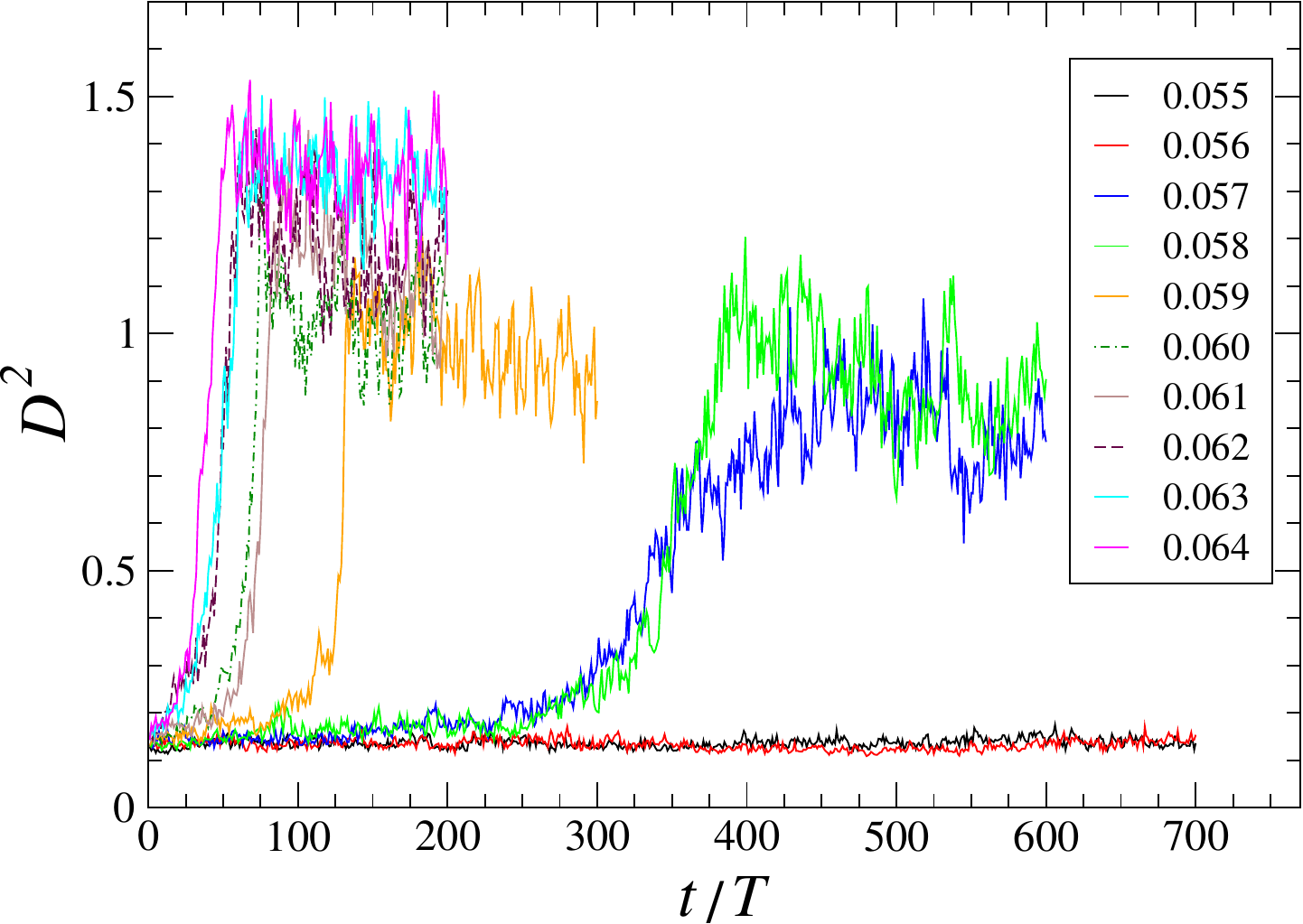}
\caption{(Color online) The average of the nonaffine measure,
$D^2[(n-1)\,T,T]$ (in units of $\AA^2$), versus number of cycles for
the indicated strain amplitudes. The period of deformation is
$T=1.0\,\text{ns}$. }
\label{fig:d2min_ave_ncyc_amp}
\end{figure}

In Fig.\,\ref{fig:d2min_ave_ncyc_amp}, we plot the nonaffine
measure, $D^2[(n-1)\,T,T]$, as a function of the cycle number $n$
for strain amplitudes $0.055 \leqslant \gamma_0 \leqslant 0.064$.
The nonaffine quantity, Eq.\,(\ref{Eq:D2min}), was evaluated for
atomic configurations at the beginning and end of each cycle when
strain is zero and then averaged over all atoms.  A close comparison
of the data for the potential energy in
Fig.\,\ref{fig:poten_T300_r10e10_amps} and the nonaffine measure in
Fig.\,\ref{fig:d2min_ave_ncyc_amp} reveals a very similar functional
dependence of $U$ and $D^2$ on the cycle number during the yielding
transition for each $\gamma_0$. Notice, for example, that the number
of cycles until yielding at $U=-4.948\,\text{eV/atom}$ and
$D^2=0.5\,\AA^2$ are essentially the same for strain amplitudes
$0.057 \leqslant \gamma_0 \leqslant 0.064$. These results indicate
that the potential energy of atomic configurations that evolve via
plastic deformation under periodic shear strongly correlates with
the average nonaffine displacements of atoms during each cycle.

\vskip 0.05in

\begin{figure}[t]
\includegraphics[width=12.0cm,angle=0]{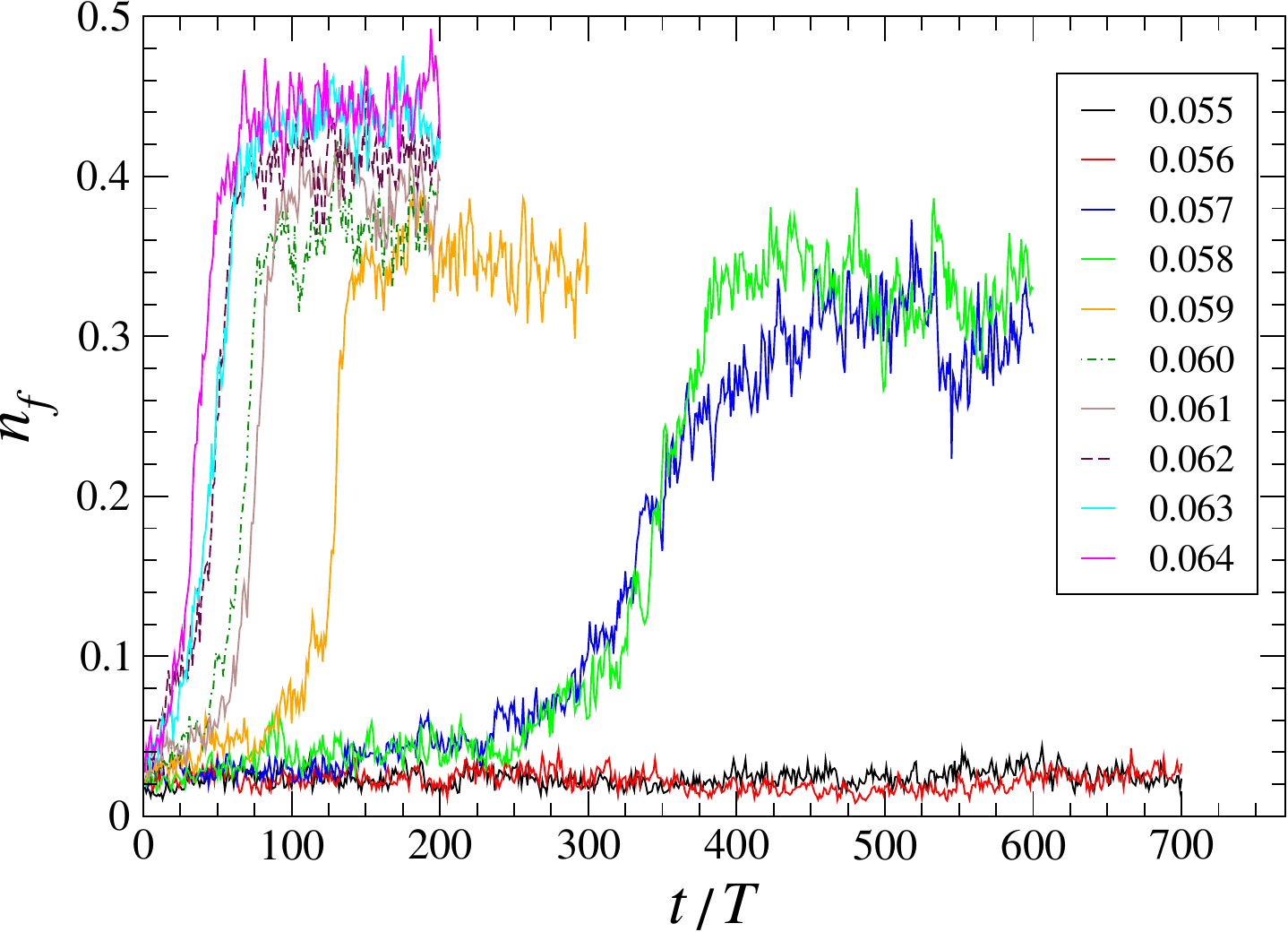}
\caption{(Color online) The fraction of atoms with large nonaffine
displacements during one cycle, $D^2[(n-1)\,T,T]>0.49\,\AA^2$, as a
function of the cycle number, $n=t/T$, for the indicated strain
amplitudes $\gamma_0$. The oscillation period is $T=1.0\,\text{ns}$.
}
\label{fig:nf_d2min_gt049_ncyc_amp}
\end{figure}

A more direct way to account for plastic events is to compute a
fraction of atoms whose nonaffine displacements during one cycle are
slightly greater than the typical cage size. In
Fig.\,\ref{fig:nf_d2min_gt049_ncyc_amp}, the fraction of atoms with
relatively large nonaffine displacements,
$D^2[(n-1)\,T,T]>0.49\,\AA^2$, is shown as a function of the cycle
number for the same strain amplitudes as in
Fig.\,\ref{fig:d2min_ave_ncyc_amp}. It can be observed in
Fig.\,\ref{fig:nf_d2min_gt049_ncyc_amp} that the fraction of atoms
varies from about 0.02 in the elastic regime at $\gamma_0=0.055$ and
$0.056$ to about 0.45 when a shear band is formed at large strain
amplitudes $\gamma_0=0.063$ and $0.064$. As is evident, the function
$n_f(n)$ clearly shows an accumulation of plastic events followed by
the yielding transition and extended plastic flow for $\gamma_0 >
0.056$.

\vskip 0.05in

\begin{figure}[t]
\includegraphics[width=12.0cm,angle=0]{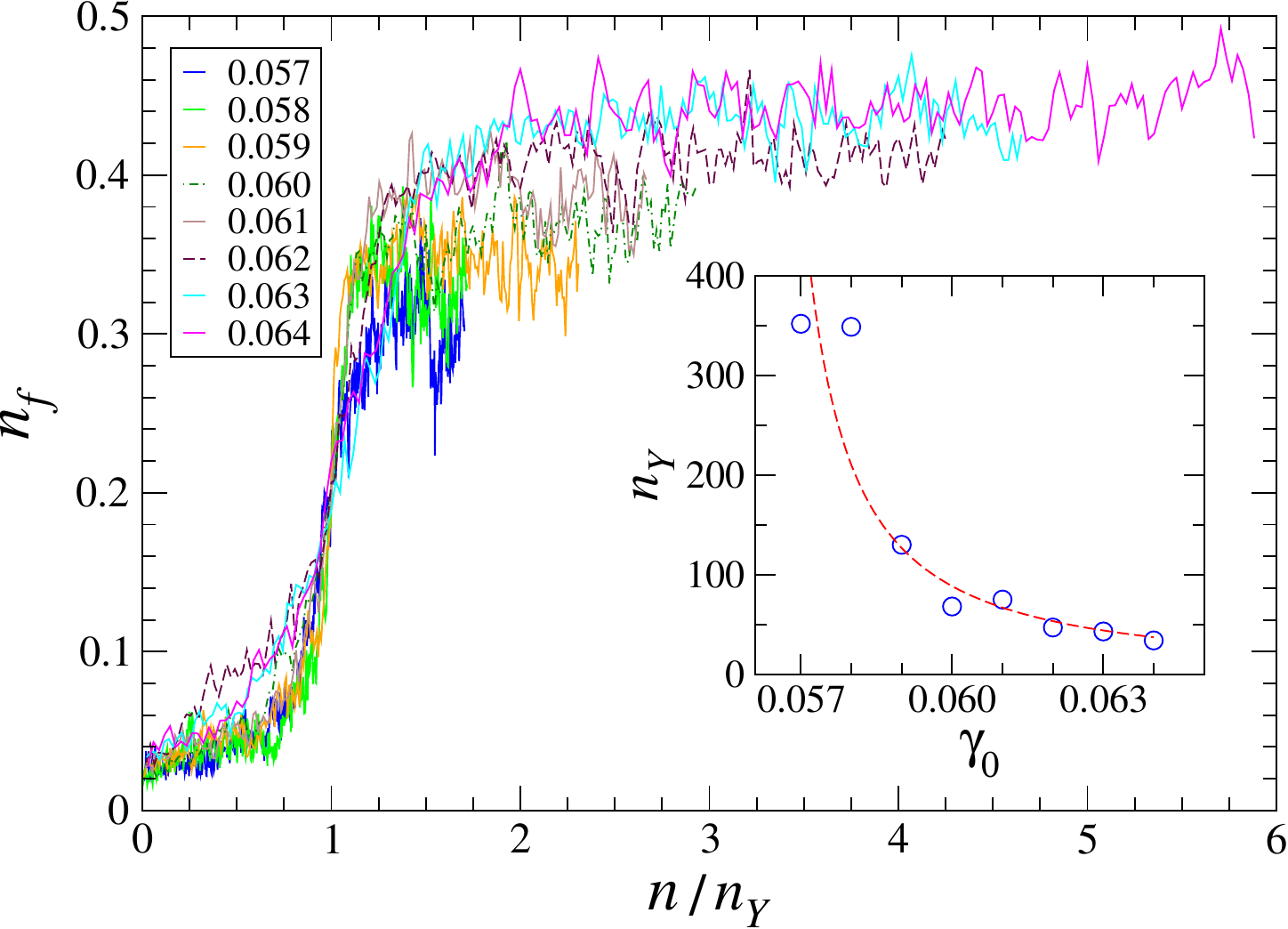}
\caption{(Color online) The fraction of atoms with large nonaffine
displacements during one cycle, $D^2[(n-1)\,T,T]>0.49\,\AA^2$,
versus $n/n_Y$, where $n=t/T$ and $n_Y$ is the number of cycles
until yielding. The inset shows $n_Y$ as a function of $\gamma_0$.
The dashed line is the best fit to the data given by
Eq.\,(\ref{Eq:fit_nY}). }
\label{fig:nf_d2min_gt049_ncyc_amp_scaled}
\end{figure}

The same data for the fraction $n_f$ are replotted in
Fig.\,\ref{fig:nf_d2min_gt049_ncyc_amp_scaled} as a function of the
the ratio $n/n_Y$, where $n=t/T$ is the cycle number and $n_Y$ is
the number of cycles to reach the yielding transition. Although the
data are somewhat noisy, it can be clearly seen that all curves
approximately follow the same trend. In our analysis, the values of
$n_Y$ were computed at $n_f=0.2$ for $\gamma_0 > 0.056$ and reported
in the inset to Fig.\,\ref{fig:nf_d2min_gt049_ncyc_amp_scaled}. The
dependence of the number of cycles until yielding on the strain
amplitude can be well captured by the power-law function:
\begin{equation}
n_Y(\gamma_0)=0.089\cdot(\gamma_0-\gamma_\text{c})^{-1.25},
\label{Eq:fit_nY}
\end{equation}
where the critical strain amplitude is taken to be
$\gamma_\text{c}=0.056$. The best fit to the data yields an exponent
of $-1.25$ in Eq.\,(\ref{Eq:fit_nY}), implying a possible divergence
of $n_Y$ upon approaching $\gamma_\text{c}$ from above. In practice,
the power-law function, Eq.\,(\ref{Eq:fit_nY}), can be used not only
to determine the number of cycles to failure at a given strain
amplitude but also to estimate a number cycles needed to rejuvenate
the glass by inducing small-scale plastic events. For example, when
the glass is loaded for $n=n_{Y}/2$ cycles, then the fraction of
plastic displacements becomes $n_f\approx 0.05$ for $\gamma_0 >
0.056$ (see Fig.\,\ref{fig:nf_d2min_gt049_ncyc_amp_scaled}).

\vskip 0.05in


Note that a power-law dependence with a different exponent of
$-1.66$ was reported for periodically sheared binary LJ glasses at a
temperature well below $T_g$~\cite{PriezCMS23}. It should be
commented that the power-law function, $n_Y(\gamma_0)$, describes
the data for LJ and Cu-Zr glasses that were both prepared by cooling
across the glass transition at a `computationally' slow rate.
However, it was found that the critical strain amplitude at zero
temperature increases when a glass is relocated into deeper regions
of its energy landscape by using either the swap Monte Carlo
algorithm~\cite{KawBer20} or mechanical
annealing~\cite{BhaSastry21}. A recent study showed an example of
how the yielding transition becomes increasingly delayed in more
stable (mechanically annealed) glasses at low temperatures when
cyclic loading is applied at a fixed strain
amplitude~\cite{Priez20delay}. Moreover, MD simulations of
periodically loaded LJ glasses at about half $T_g$ revealed that
although the yielding transition is delayed in better annealed
glasses, the critical strain amplitude remains
unchanged~\cite{PriezJNCS22}. Therefore, it is still unclear how, in
general, the critical strain amplitude and the power-law dependence,
$n_Y(\gamma_0)$, are affected by the degree of annealing for
cyclically loaded glasses at a finite temperature.

\vskip 0.05in

\begin{figure}[t]
\includegraphics[width=12.0cm,angle=0]{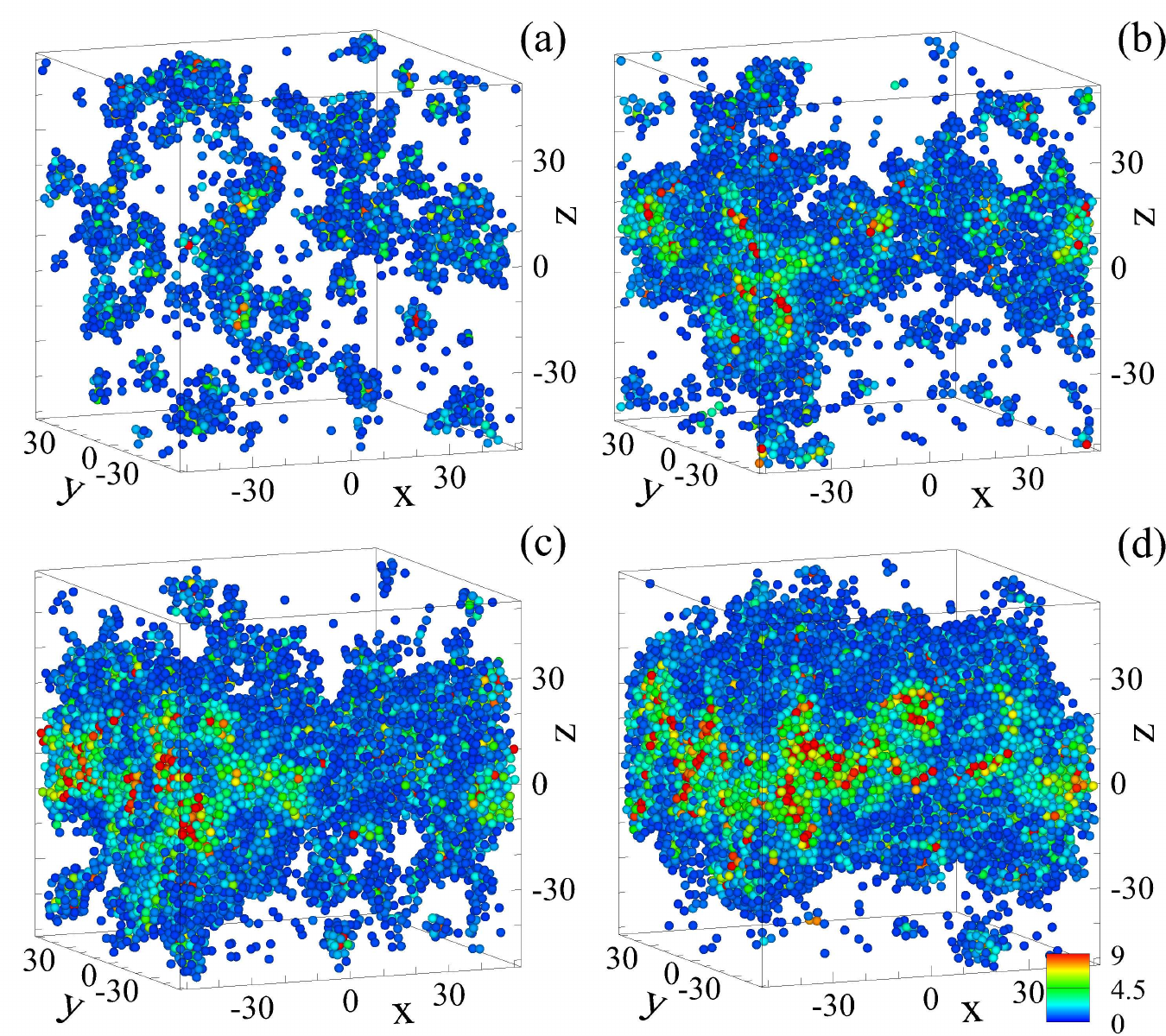}
\caption{(Color online) The atomic configurations of the metallic
glass periodically loaded at the strain amplitude $\gamma_0=0.064$.
The nonaffine displacements are only shown for atoms with the
nonaffine measure (a) $D^2(3\,T,T)>0.49\,\AA^2$, (b)
$D^2(30\,T,T)>0.49\,\AA^2$, (c) $D^2(40\,T,T)>0.49\,\AA^2$, and (d)
$D^2(80\,T,T)>0.49\,\AA^2$. The magnitude of $D^2$ is indicated by
the legend color. The Zr and Cu atoms are not depicted to scale. }
\label{fig:snapshot_amp064}
\end{figure}
\begin{figure}[t]
\includegraphics[width=12.0cm,angle=0]{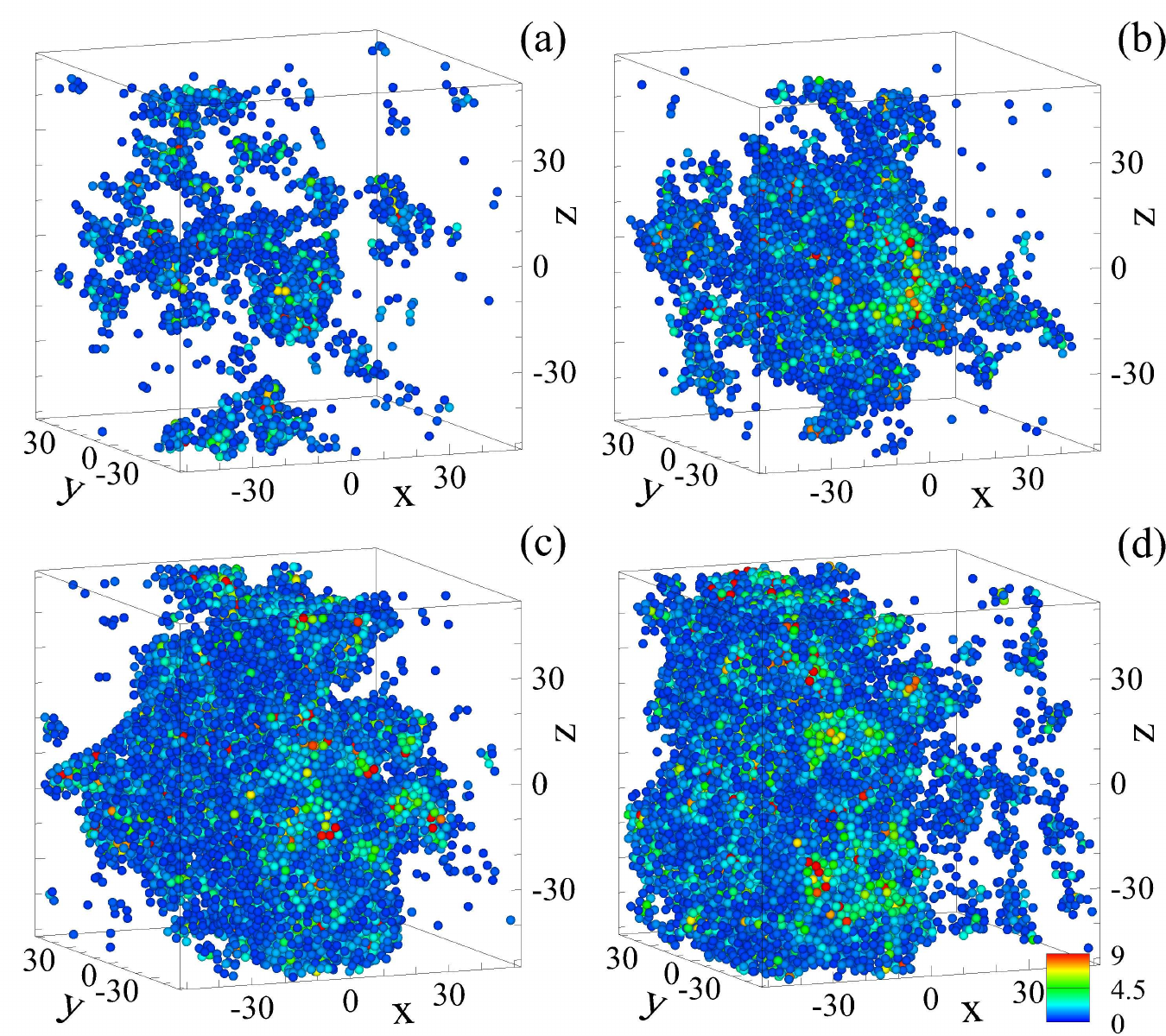}
\caption{(Color online) Selected configurations of atoms in the
binary glass deformed with the strain amplitude $\gamma_0=0.057$.
The nonaffine measure is (a) $D^2(150\,T,T)>0.49\,\AA^2$, (b)
$D^2(320\,T,T)>0.49\,\AA^2$, (c) $D^2(380\,T,T)>0.49\,\AA^2$, and
(d) $D^2(500\,T,T)>0.49\,\AA^2$. The magnitude of $D^2$ is indicated
according to the legend in the panel (d).  }
\label{fig:snapshot_amp057}
\end{figure}

The spatial extent of plastic events in the metallic glass under
cyclic shear is illustrated in Fig.\,\ref{fig:snapshot_amp064} for
the strain amplitude $\gamma_0=0.064$ and in
Fig.\,\ref{fig:snapshot_amp057} for $\gamma_0=0.057$. For clarity,
only atoms with large nonaffine displacements during one cycle,
$D^2[(n-1)\,T,T]>0.49\,\AA^2$, are shown when strain is zero. It can
be seen in Fig.\,\ref{fig:snapshot_amp064}\,(a) that during the
fourth cycle at the strain amplitude $\gamma_0=0.064$, atoms with
large nonaffine displacements form several isolated clusters. Upon
further loading, the steep increase in $U$ (see
Fig.\,\ref{fig:poten_T300_r10e10_amps_057_064}) and $n_f$ (see
Fig.\,\ref{fig:nf_d2min_gt049_ncyc_amp}) is characterized by the
formation of an extended cluster of atoms that percolates across the
entire system, as shown in Fig.\,\ref{fig:snapshot_amp064}\,(b, c).
After 80 cycles, the shear band develops along the $xy$ plane, shown
in Fig.\,\ref{fig:snapshot_amp064}\,(d), which is consistent with
the plateau level in the potential energy reported in
Fig.\,\ref{fig:poten_T300_r10e10_amps_057_064}. A similar trend can
be observed in Fig.\,\ref{fig:snapshot_amp057} for the strain
amplitude $\gamma_0=0.057$, except that the yielding transition is
delayed by about 300 cycles and the shear band is formed along the
perpendicular plane $yz$. The results in
Fig.\,\ref{fig:snapshot_amp057}\,(a) also confirm that after about
$n_Y/2$ cycles at $\gamma_0=0.057$, plastic deformation involves
only small-scale clusters ($n_f\approx0.04$).

\section{Conclusions}

In summary, we studied the effect of periodic shear deformation on
the yielding transition in metallic glasses using molecular dynamics
simulations. We considered a $\text{Cu}_{50}\text{Zr}_{50}$ metallic
glass prepared by slow cooling from the melt to room temperature and
then subjected to oscillatory shear at constant volume. It was shown
that during loading for hundreds of cycles at strain amplitudes
slightly below a critical value, the potential energy at the end of
each cycle remains nearly unchanged and plastic deformation proceeds
via localized clusters of atoms with relatively large nonaffine
displacements. When loaded at larger strain amplitudes, the glass
becomes gradually rejuvenated via collective irreversible
displacements of atoms from cycle to cycle until the yielding
transition and the formation of a system-spanning shear band.
Remarkably, the number of cycles to reach the yielding transition
approximately follows a power-law dependence on the difference
between the strain amplitude and the critical value. Despite a
different value of the power-law exponent, these conclusions are in
agreement with the results for binary Lennard-Jones glasses
periodically deformed at a temperature well below the glass
transition temperature.

\section*{Acknowledgments}

Financial support from the National Science Foundation (CNS-1531923)
is gratefully acknowledged. Molecular dynamics simulations were
performed at the Wright State University's Computing Facility and
the Ohio Supercomputer Center using the LAMMPS code~\cite{Lammps}.



%

%

%

%

%

%

%

%

\bibliographystyle{prsty}

\end{document}